\documentclass[conference,a4paper]{IEEEtran}
\usepackage[singlespacing]{setspace} 
\IEEEoverridecommandlockouts
\usepackage{cite}
\usepackage{amsmath,amssymb,amsfonts}
\usepackage{algorithm}
\usepackage{algpseudocode}
\usepackage{graphicx}
\usepackage{textcomp}
\usepackage{blindtext}
\usepackage{subfigure}
\usepackage{float} 
\usepackage{subfigure}
\usepackage{xcolor}
\usepackage{diagbox}
\usepackage{slashbox}
\usepackage{wrapfig}
\usepackage{array}
\usepackage{makecell}
\usepackage{multirow}

 \usepackage[left=1.7cm,right=1.7cm,top=1.9cm]{geometry}
\def\BibTeX{{\rm B\kern-.05em{\sc i\kern-.025em b}\kern-.08em T\kern-.1667em\lower.7ex\hbox{E}\kern-.125emX}}
\makeatletter

\begin{document}
	\title{A Survey on the Role of Artificial Intelligence and Machine Learning in 6G-V2X Applications}
	\IEEEpeerreviewmaketitle
	\author{\IEEEauthorblockN{Donglin Wang\IEEEauthorrefmark{2}, Anjie Qiu\IEEEauthorrefmark{2}, Qiuheng Zhou\IEEEauthorrefmark{1}, and Hans D. Schotten\IEEEauthorrefmark{2}\IEEEauthorrefmark{1}}
		\IEEEauthorblockA{\textit{\IEEEauthorrefmark{2}Rhineland-Palatinate Technical University of Kaiserslautern-Landau, Germany} \\
		$\{$dwang,qiu,schotten$\}$@eit.uni-kl.de \\}
		\IEEEauthorblockA{\textit{\IEEEauthorrefmark{1}German Research Center for Artificial Intelligence (DFKI GmbH), Kaiserslautern, Germany} \\
		$\{$qiuheng.zhou,schotten$\}$@dfki.de}
	}
	\maketitle
\begin{abstract}
The rapid advancement of Vehicle-to-Everything (V2X) communication is transforming Intelligent Transportation Systems (ITS), with 6G networks expected to provide ultra-reliable, low-latency, and high-capacity connectivity for Connected and Autonomous Vehicles (CAVs). 
Artificial Intelligence (AI) and Machine Learning (ML) have emerged as key enablers in optimizing V2X communication by enhancing network management, predictive analytics, security, and cooperative driving due to their outstanding performance across various domains, such as natural language processing and computer vision. This survey comprehensively reviews recent advances in AI and ML models applied to 6G-V2X communication. It focuses on state-of-the-art techniques, including Deep Learning (DL), Reinforcement Learning (RL), Generative Learning (GL), and Federated Learning (FL), with particular emphasis on developments from the past two years. Notably, AI, especially GL, has shown remarkable progress and emerging potential in enhancing the performance, adaptability, and intelligence of 6G-V2X systems. Despite these advances, a systematic summary of recent research efforts in this area remains lacking, which this survey aims to address. We analyze their roles in 6G-V2X applications such as intelligent resource allocation, beamforming, intelligent traffic management, and security management. Furthermore, we explore the technical challenges, including computational complexity, data privacy, and real-time decision-making constraints, while identifying future research directions for AI-driven 6G-V2X development. This study aims to provide valuable insights for researchers, engineers, and policymakers working towards realizing intelligent, AI-powered V2X ecosystems in 6G communication.

% This survey presents a comprehensive review of AI and ML models in 6G-enabled V2X communication, highlighting cutting-edge techniques such as Deep Learning (DL), Reinforcement Learning (RL), Generative Learning (GL), and Federated Learning (FL) from the last two years. In the past two years, AI, particularly GL, has made significant advancements and demonstrated considerable potential for 6G-V2X communication in many newly research works, which haven't been summarized. 

\end{abstract}


\section{Introduction}
The upcoming 6G of wireless networks is poised to revolutionize vehicular communication, commonly known as Vehicle-to-Everything (V2X), by offering unprecedented levels of performance and reliability \cite{noor20226g}. V2X communication has garnered significant research attention from both academia and industry. As a crucial enabler of Intelligent Transportation Systems (ITS), V2X integrates various wireless technologies, including Vehicle-to-Vehicle (V2V), Vehicle-to-Infrastructure (V2I), and Vehicle-to-Pedestrian (V2P) communications. It also facilitates interactions with Vulnerable Road Users (VRUs) and cloud Networks (V2N) [1]. Vehicle-to-Satellite (V2S) \cite{3GP15} is gaining attention as a key component of future 6G vehicular networks, particularly for enhancing connectivity in remote or underserved areas. 

The two dominant technologies in V2X communication are: 1. Dedicated Short-Range Communication (DSRC), which is based on the IEEE 802.11p standard \cite{IEEE07}. 2.  Cellular-V2X (C-V2X), which operates on 4G/LTE \cite{3GP16}, 5G networks \cite{3GP19} and the coming 6G \cite{3GP21}. The overarching vision is that the 6G of wireless systems will empower V2X communications. The development of 6G-V2X communication is shaping the future of connected mobility. It aims to address the limitations of current technologies like 5G by offering ultra-high data rates, sub-millisecond latency, and enhanced energy efficiency. 

In recent decades, Artificial Intelligence (AI) and Machine Learning (ML) have gained significant attention due to their exceptional performance in various fields, such as natural language processing and computer vision \cite{nokia2023} \cite{6gflagship}. With ongoing standardization efforts to integrate AI and ML into beyond 5G and 6G networks, these technologies are promising to optimize user experience, enhance network control and management, and advance applications in road safety and entertainment. 

Last several years, numerous studies have been conducted to explore the application of AI and ML technologies in the development of 6G-V2X communication systems. \cite{sanghvi2021res6edge} introduces Res6Edge, a scheme designed to improve C-V2X communication by leveraging 6G network capabilities and Edge AI (Edge-AI). It aims to overcome the limitations of previous 5G-based approaches, particularly in terms of resource sharing, latency, and reliability in the context of connected and Autonomous Vehicles (CAVs). The core idea is to utilize the advanced features of 6G, such as higher bandwidth and lower latency, in conjunction with AI processing at the edge of the network. This enables faster decision-making, real-time resource allocation, and improved learning for cooperative sensing and communication among CAVs. The authors propose a three-phase approach, the layered network model, 6G resource allocation, and an intelligent Edge-AI scheme. \cite{noor20226g} provides a comprehensive overview of potential technologies to shape future 6G-V2X communications, highlighting key advancements in materials, algorithms, and system architectures. ML is emphasized for its instrumental role in advancing vehicular communication and networking, particularly in enabling intelligent transportation systems. \cite{tong2019ai} and  \cite{christopoulou2023artificial} both conduct surveys of AI enabled V2X communication. This paper \cite{tong2019ai} surveys the use of AI to address research challenges in V2X systems, highlighting AI's potential to optimize traditional data-driven approaches for intelligent traffic solutions. It summarizes research contributions, categorizing them by application domains within V2X, including traffic efficiency, road safety, and energy efficiency. The paper identifies open problems and research challenges that need to be addressed to fully realize AI's potential in advancing V2X systems, such as consumer trust, privacy, and balancing fairness and optimization in traffic light control. In this paper \cite{christopoulou2023artificial}, it provides a thorough examination of the intersection between AI, ML, and V2X communications. The authors discuss the critical role of these technologies in optimizing various operations within vehicular networks, which are essential for advancing autonomous driving and ensuring road safety. The paper presents a systematic review of the literature focusing on the integration of AI/ML within V2X communications and categorizes their exploration into several prominent operational areas like handover management, physical resource allocation, and Quality of Service (QoS) prediction etc. The authors analyze existing research papers to extract AI/ML techniques, training features, architectural configurations (centralized, distributed, federated), and operational challenges that arise in dynamic vehicular environments, which add substantial complexity to ML tasks. However, some of the research papers referenced in both work are outdated due to the rapid evolution of AI technologies, and certain emerging applications of 6G-V2X communication are not covered in this study. 

In this work, we conduct a comprehensive review of recent advancements in AI for 6G-V2X communication, especially in 2023-2025, and identify several representative and widely adopted AI models explicitly suited for 6G-V2X networks. Our selection focuses on models that demonstrate high performance, adaptability, and effectiveness in addressing key challenges in 6G-V2X, such as real-time decision-making, resource optimization, and network reliability.

 % The convergence of Large Language Models (LLMs) \cite{zhao2023llm}, particularly Generalized Pretrained Transformers (GPT) \cite{achiam2023gpt}, with pregenerative AI and traditional ML algorithms represents a paradigm shift from static, rule-based systems to adaptive, learning-driven frameworks. This integration leverages both established methodologies and emerging AI innovations to enhance the efficiency and intelligence of next-generation networks \cite{LLMwhitepaper2025}. While LLMs offer substantial potential within mobile and 6G networks, they will function in conjunction with conventional AI/ML models to ensure robust and scalable solutions.

\section{representative AI models for V2X communcation}
\begin{figure}[htbp]
	\centering
	\includegraphics[width=\linewidth]{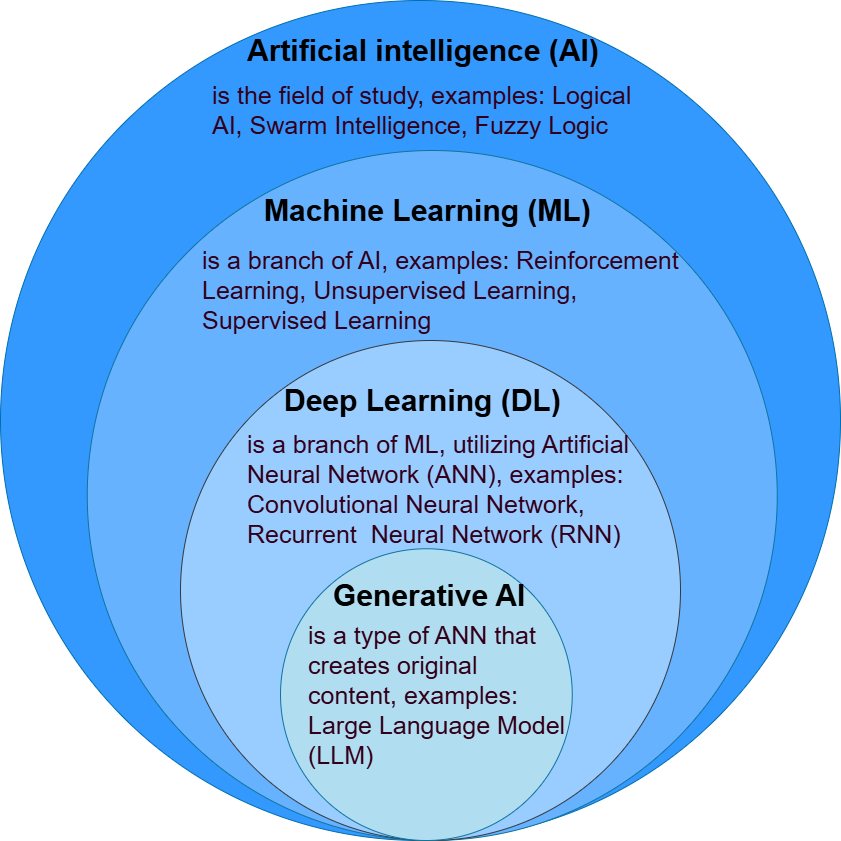}
	\caption{AI Hierarchy}
	\label{AI}
\end{figure}

Recent advancements in computing technology and hardware have made AI a fundamental component in almost every engineering research domain. Particularly through ML, it has been effectively integrated into various domains, including monitoring the orbital environment as deep-space exploration has evolved. While AI systems, such as Artificial Neural Networks (ANN), have shown potential in enhancing decision-making processes, particularly in road transportation contexts\cite{AJ:olayode2020}, one notable application is autonomous driving, where AI is essential to reproduce key aspects of human driving \cite{tong2019ai}. 
Numerous AI models are designed for various applications, each tailored to specific domains as shown in Figure \ref{AI}. This diagram shows the hierarchy in AI technologies where AI is the broadest field focused on simulating human intelligence; ML is a subset of AI that enables systems to learn from data; DL is a type of ML using neural networks for complex tasks, and Generative AI is a form of DL that creates new content, like text or images.

\subsection{Machine Learning (ML)}

ML, as a central subfield of AI, focuses on how these agents can improve their perception based on experience or data \cite{AJ:omar2021}. ML algorithms emphasize the ability of machines to improve their performance on a task over time without being explicitly programmed for each specific task. Recently, ML models have progressed rapidly from traditional supervised and unsupervised techniques like Reinforcement Learning (RL) to DL and, currently, to Generative Pre-trained Transformers (GPT) for processing natural language. Furthermore, the architecture of ML is shifting from centralized systems to distributed frameworks, including Federated Learning (FL) \cite{ali2021machine}. RL is applied to optimize decision-making processes in dynamic environments, such as real-time traffic management and adaptive routing \cite{gyawali2021deep}. Agents learn to make optimal decisions based on feedback from their environment, improving network operations over time. This approach is highly adaptable to changing conditions, making it suitable for V2X applications that require real-time responses, such as managing handovers and resource allocation in vehicular networks. However, the exploration phase can be resource-intensive, requiring significant time and interactions to learn effective policies.
\par
This paper \cite{tan2022machine} surveys ML applications in vehicular networks, highlighting network control, resource management, and energy efficiency. It identifies challenges like network forming/deforming decisions and advocates for multi-agent cooperative methods and complexity reduction using mobile edge computing.

% When discussing ML models in the 6G-V2X systems, the agent-based learning is often discussed, while the agent usually refers to moving vehicles or the Road-side Units (RSUs), interacting with their environment by receiving messages of state representations, selecting actions by decisions, receiving rewards according to the environment, allowing them to learn and optimize their actions over time.

\subsection{Deep Learning (DL)}

DL \cite{aggarwal2018dl} is a crucial branch of ML, as shown in Figure \ref{AI}, that significantly outperforms traditional ML techniques, especially when large amounts of data are available. Unlike conventional approaches that rely on manual feature extraction, DL leverages ANNs \cite{yao1999ann} to automatically learn complex patterns and representations from raw data. ANNs are inspired by the structure and function of the human nervous system, consisting of interconnected layers of artificial neurons that hierarchically process information. This ability to learn intricate relationships makes DL particularly effective in tasks such as resource allocation and management, security, safety, risk prediction, and autonomous driving and network management in V2X networks, etc.. DL also refers to Deep Reinforcement Learning (DRL) and is introduced as a method in the 6G-V2X systems where Neural Networks (NNs) approximate the expected value of state-action pairs, enhancing the agent's ability to learn from trial and error in environments with large state spaces \cite{marzuk2025deep}. 
\par
This survey paper \cite{aradi2020survey} examines the use of DRL for motion planning in autonomous vehicles, focusing on strategic decision-making, trajectory planning, and control. It reviews environment modeling, state representations, rewarding mechanisms, and NN implementations crucial for DRL system design, while also covering vehicle models, simulation platforms, and computational needs. The study categorizes state-of-the-art DRL solutions based on autonomous driving tasks like car-following, lane-keeping, merging, and driving in dense traffic, and it concludes by discussing open challenges and future research directions in the field.

% Waqas et al. \cite{AJ:Syed_Muhammad_Waqas_2024} use Flexible Graph Neural Networks (FGNNs)  to optimize the resource distribution problem in V2X networks based on instantaneous channel conditions. Hu et al.\cite{AJ:Bintao_Hu_2024} introduce an algorithm with Multi-Agent Deep Deterministic Policy Gradient (MADDPG) to compute offloading and resource allocation in ISAC-aided 6G-V2X networks. Ahmed et al. \cite{AJ:Tahir_H_Ahmed_2023} deploys an Enhanced Deep Cooperative Q-Learning (DCO-DQN) to incorporate an advance reward function to reflect multiple performance metrics, ensuring robust performance across volatile wirelsses channels among moving vehicles. Park et al. \cite{AJ:SeungYoung_Park_2024} uses combined Rule-based and DL-based misbehavior detection to leverage Multi-access Edge Computing (MEC) to decentralize processing, reducing latency and improving detection accuracy among the vehicular network. Ding et al. \cite{AJ:Ding} introduce a DRL to aid V2X network about lane-changing for emergency vehicle preemption in connected autonomous transport systems.

\subsection{Generative Learning (GL)}

% Generative Learning (GL) is a new generation of ML approach where models are trained to generate new data instances that resemble the training data, especially in the domain of natural language, visual context, and image generation. 

Generative Learning (GL) typically refers to models that learn the joint probability distribution of the data, like Generative Adversarial Networks (GANs) \cite{gui2021gans}, Variational Autoencoders (VAEs) \cite{doersch2016tutorial}, etc. These models can generate new samples similar to the training data, e.g., images, text  \cite{zhang2024generative}. Large Language Models (LLMs)\cite{zhao2023llm} is also one of the most representative GL models, which is designed to understand and produce human language \cite{AJ:Zhiyuan_Zhou_2024}. These models are trained on vast amounts of text data, allowing them to generate coherent and contextually appropriate text based on the input they receive. The emergent capabilities of LLMs, such as instruction following and in-context learning, highlight their generative nature, as they can create responses that are not merely regurgitation of the training data but rather novel outputs based on learned patterns. LLMs are advanced AI systems, like GPT-4 \cite{achiam2023gpt}, that use transformer-based NN trained on massive text datasets to understand, generate, and manipulate human language for tasks like writing, translation, or reasoning. LLMs and GPTs can play a crucial role, especially as we tackle the demands of 6G-V2X communication, a field where ensuring connectivity, capacity, latency, mobility, and reliability becomes increasingly difficult without AI \cite{tataria20216g}. 

In this paper \cite{tong2024connectgpt}, Kailin Tong and Selim Solmaz present an innovative framework for enhancing the functionality of CAVs by merging capabilities of LLMs, specifically leveraging GPT-4, with vehicular communication technologies, particularly V2X communication. The primary focus is on automating the generation of standardized safety messages within dynamic traffic environments. The methodology developed in this paper involves a multi-step pipeline that integrates data flow from infrastructure sensors, primarily using advanced camera systems, into a processing framework featuring the GPT-4 model for interpreting traffic conditions and generating communications. This paper \cite{panyam2025survey} surveys the integration of LLMs and AI agents into V2X communication systems, focusing on their application in simulation, analysis, and architectural design. It explores how these AI technologies enhance realistic scenario generation, adaptive data processing, and real-time decision-making within V2X infrastructures, incorporating interdisciplinary perspectives like cloud computing and data engineering. The survey identifies emerging trends and future research directions, such as scalable deployment, privacy preservation, and standardization, highlighting the transformative potential of AI in creating smarter and safer transportation systems. Using LLM for 6G-V2X communication is an emerging and innovative concept that combines natural language processing NLP with intelligent transportation systems. 
% Masri et al \cite{AJ:Sari_Masri_2025} uses GPT-4 for demonstrating high accuracy in conflict identification and decision-making in reliable autonomous driving as a traffic control aider. Yan et al \cite{AJ:Zijiang_Yan_2024} combines LLMs with Double Deep Q-Learning (DDQN) to optimize the V2I communication and autonomous driving policies, results in reduction of handovers and faster convergence to higher rewards compared to conventional DDQN algorithms. Zhou et al \cite{AJ:Zhiyuan_Zhou_2024} uses LLMs to integrate decision making frameworks for autonomous vehicles, enabling context-aware and human-like driving behaviors, which quantifies risks and refine decisions through iterative learning against real-world datasets. David et al \cite{AJ:Teef_David_2024} deploys generative AI to optimize traffic management bz simulating complex network scenarios and adaptive communication schemes, leading to groundbreaking improvements in vehicular network efficiency and safety.
\subsection{Federated Learning (FL)}
%draw a diagram
Unlike other ML models we mentioned before, FL \cite{samarakoon2019fl} is a distributed ML setting where multiple entities collaborate to solve a learning problem, without directly exchanging data. Training ML models at base stations or vehicles is crucial for efficient 6G-V2X applications. While cloud-based training is common, it can be slow due to dynamic network conditions and high data transfer costs from the edge. Local training is preferred, but limited data at individual nodes can impact model accuracy. Joint training offers a solution, though privacy concerns limit data sharing. FL addresses both privacy and communication overhead, making it a promising approach for wireless networks \cite{noor20226g}. FL in 6G-V2X can be implemented using either centralized or decentralized architectures. Centralized FL relies on a central server to aggregate local models from vehicles, while decentralized FL distributes the aggregation process across vehicles and edge nodes. Decentralized architectures are more robust against single-point failures but face challenges in scalability and communication overhead\cite{AJ:Biagio_Boi_2024}. 

% In V2X environments, vehicles train local models using their onboard data and periodically upload model updates to a central server or edge nodes. The aggregation process combines these updates to form a global model, which is then disseminated back to the vehicles. Techniques such as asynchronous federated learning and split learning have been proposed to address the challenges of intermittent connectivity and computational constraints in resource-constrained vehicles\cite{AJ:Levente_Alekszejenko_2024}.

This survey \cite{zhang2024survey} reviews the application of FL in ITS, focusing on traffic flow prediction, traffic target recognition, and vehicular edge computing. FL addresses limitations of centralized training by enabling decentralized model training, preserving data privacy, and improving real-time performance in ITS scenarios. The paper also explores future research directions for FL-based ITS, including space-air-ground integrated FL, LLMs, and ethical/privacy compliance.

% \par
% Bhardwaj et al \cite{AJ:Sanjay_Bhardwaj_2024} uses FL to maximize the throughput among users and optimize the uplink resources between cellular and vehicular users. Qiang et al \cite{AJ:Xing_Qiang_2024} apply Adaptive Split Federated Learning (ASFV) to address vehicle heterogeneity to dynamic network conditions, by adaptively splitting models and parallelizing the training process. 

% In the domain of autonoumous driving, baucas, Baucas et al \cite{AJ:Marc_Jayson_Baucas_2024} deploy an FL architecture to secure object recognition in CAVs while preserving data privacy from vehicular users. Devarajan et al \cite{AJ:Ganesh_Gopal_Devarajan_2025} proposes that an explainable FL deployed in transparent object detection models results in reduction of data privacy risks by 40\%. In traffic magement and predictive analytics, Fl also outperformans than the traditional ML architectures. Alekszejenk'o \& Dobrowiecki \cite{AJ:Levente_Alekszejenko_2024} mitigate congestion at intersections and optimize the traffic flow by applying FL system to provide real-time data to vehicles with predicted transportation network states.

% \begin{figure}[htbp]
% 	\centering
% 	\includegraphics[width=\linewidth]{federated-learning-model.png}
% 	\caption{Federated learning for vehicular communication}
% 	\label{fl}
% \end{figure} 

%%%%%%%%%%%%%%%%%%%%%% AI for 6G-V2X Applications %%%%%%%%%%%%%%%%%%%%%%%%%%%%%%%%%%%%%%%%%

%%requirements%%
% 6G-V2X application characteristics? latency? accuracy? real-time?
% what ML can be applied, why?
% examples, paper work, and results

\section{AI for 6G-V2X Applications}

Integrating AI, ML, DL, LLM, or FL is essential for enhancing the performance, reliability, and intelligence of 6G-V2X communication. These technologies enable real-time decision-making, predictive analytics, and automation in CAV ecosystems. In the following sections, we explore several AI-enabled applications within the context of 6G-V2X communication, highlighting their potential to address key challenges and enhance the performance and reliability of next-generation vehicular networks as summarized in Table \ref{tab:6g-v2x-ai}.

%%resource allocation
%%find researches on AI for resource allocations 
\subsection{Intelligent Resource Management} 

Resource allocation within V2X communication systems is a foundational component in developing and advancing ITS and related applications. As 6G-V2X communication networks become increasingly dynamic, heterogeneous, and performance-sensitive, the design of intelligent and adaptive resource allocation strategies becomes imperative. In recent studies, AI has been extensively explored as a powerful tool for resource allocation in 6G-V2X communication systems. A wide range of novel methodologies have been proposed and employed to enhance the effectiveness, efficiency, and adaptability of resource allocation mechanisms, addressing the complex and dynamic requirements characteristic of next-generation vehicular networks.
\par
RL \cite{he2017rl} plays a crucial role in optimizing spectrum sharing, bandwidth allocation, and network slicing in 6G-V2X communication by enabling intelligent, autonomous decision-making in dynamic network environments. \cite{ji2024rlra} and \cite{khan2024rlra} employ DRL techniques for resource allocation. In \cite{ji2024rlra}, this paper introduces a GNN and DRL method for resource allocation in C-V2X to enhance V2V communication success while minimizing V2I interference. The proposed method constructs a dynamic graph, uses the GraphSAGE model for adaptation, and enables vehicles to independently make resource allocation decisions based on local observations with structural information. Simulation results demonstrate that the GNN-enhanced method improves decision-making quality with a modest computational increase, outperforming other methods in ensuring high V2V success rates and V2I transmission rates. This paper \cite{khan2024rlra} introduces a DRL framework for joint power and block length allocation to minimize decoding-error probability in ultra-reliable low-latency 6G-V2X communication systems. An event-triggered mechanism is integrated into the DRL framework to reduce DRL process executions while maintaining reasonable reliability performance by assessing when to initiate the DRL process. Simulation results demonstrate that the proposed event-triggered DRL scheme achieves 95\% of the performance of the joint optimization scheme while reducing DRL executions by up to 24\% for different network settings.
\par
% LLM in resource allocation 
This latest work \cite{liu2024llm} integrates LLMs into vehicles for enhanced V2X communication. It utilizes edge computing to balance local and RSU-based processing of LLM tasks. Formulating a multi-objective optimization problem decomposed into sub-problems solved via Sequential Quadratic Programming (SQP) and fractional programming to harmonize completion speed with energy efficiency. 
\par
FL is also utilized as a method to optimize resource management in the context of emerging 6G-V2X communications \cite{bhardwaj2024flrl} \cite{prathiba2021FL}. This new paper \cite{prathiba2021FL} introduces a FL empowered computation Offloading and Resource management (FLOR) framework for 6G-V2X to minimize latency by leveraging heterogeneous networks like V2X, DSRC, 5G-mmWave, and 6G-V2V. FLOR uses stochastic network calculus to estimate the upper bound delay of heterogeneous communication and Federated Q-Learning for effective Radio Resource Management, optimizing resource allocation and utilization. Simulation results demonstrate that FLOR achieves a 20.69\% improvement in computation offloading and resource allocation compared to existing solutions, enhancing the performance of autonomous vehicles. This paper \cite{bhardwaj2024flrl} introduces FL-RA-V2X, a FL-based resource allocation method for V2X communications that optimizes throughput while preserving privacy. FL-RA-V2X ensures fairness by meeting performance requirements for cellular users and outage probability constraints for vehicle users, enabling uplink resource sharing between both user types. Simulations demonstrate FL-RA-V2X's enhanced throughput efficiency, fairness, and adaptability to high mobility and asynchronous training compared to existing algorithms.

%%beamforming

\subsection{AI-Powered Beamforming in Millimeter-Wave (mmWave) \& Terahertz (THz) Bands
}

Given the anticipated widespread adoption of Millimeter-Wave (mmWave) communication \cite{mizmizi20216g} in 6G-V2X systems, beamforming and massive Multiple Input Multiple Output (MIMO) \cite{forsythe2004mimo} technologies are essential to mitigate the significant path loss challenges associated with mmWave frequencies \cite{noor20226g}. AI technologies can help optimize beam selection in mmWave and THz communication for ultra-fast and reliable V2X connections. AI-based adaptive beamforming enhances V2I and V2N communications.  
% \cite{tan2024beam} beamforming survey

The study \cite{moon2020dlbf} proposes a DL-based algorithm for channel estimation and tracking in mmWave V2X communications. For channel estimation, a Deep Neural Network (DNN) is trained to learn the mapping between received omni-beam patterns and the mmWave channel, thereby enabling the maximization of the effective achievable rate while minimizing overhead. 

Sanjay Bhardwaj et al. \cite{bhardwaj2023flbf} and \cite{bhardwaj2024flbf} have focused on applying FL techniques to beamforming in V2X communications. \cite{bhardwaj2023flbf} introduces FL-BT, a FL-based approach for joint radar and communication mmWave beamtracking in V2X environments with imperfect Channel State Information (CSI).
FL-BT uses a federated module comprised of Support Vector Regression (SVR) and an Unscented Kalman Filter (UKF) at both the vehicle user (local client) and RSU to predict future channel behavior. Simulation results demonstrate that FL-BT significantly improves V2X communication performance compared to traditional methods by effectively handling bandwidth constraints, privacy considerations, and accurately predicting future channel values. The new paper work \cite{bhardwaj2024flbf} introduces FL-mm-V2X, a FL approach for mmWave beamforming in V2X communications, addressing challenges like Doppler shift and imperfect channel state information using Non-Orthogonal Multiple Access (NOMA) and MIMO. The proposed method optimizes power allocation through iterative updates, employing Lagrangian variables and a sub-gradient approach, while also integrating a Convolutional Neural Network (CNN) for Doppler shift estimation. Simulation results demonstrate that FL-mm-V2X effectively mitigates the impact of imperfect CSI and Doppler shift, achieving lower complexity and better performance compared to existing methods in dynamic, high-mobility mmWave channels.

%%%%%%%%%%%%%%%%%%%%%%%%%%%%%%%%%%%%%%%%%%%%%%%%%%%
%Anjie Qiu 
%smart mobility application 

\subsection{Intelligent Traffic Management}

Intelligent Traffic Management (ITM) is a powerful application in 6G-V2X communication by enabling ultra-fast, low-latency, AI-driven coordination between vehicles, infrastructure, and control systems. In this work\cite{khalid2025modernizing}, it explores the domain of ITM through the lens of AI and ML. These advanced technologies present transformative solutions to longstanding challenges in transportation, such as congestion and road safety. By harnessing the power of AI and ML, smart transportation systems are developed that anticipate traffic disruptions, optimize travel efficiency, and enhance safety. In this paper \cite{qiu2024advanced}, a new scheme is proposed to generate large-scale traffic demands, where a real-world dataset is fed into an ML model to predict traffic flows. The approach is highly computationally efficient and can be applied to any city scenario. Furthermore, the training dataset and the ML model can be updated for more precise predictions. This paper \cite{zhou2024intelligent} introduces a DRL-based traffic signal control method using 6G communication technology to enhance intelligent road network management. The proposed DRL algorithm leverages real-time data to manage complex traffic flows, optimize traffic signal plans, and improve overall traffic efficiency. Simulation results demonstrate that the DRL method significantly outperforms the CNN algorithm by reducing average travel time and queue length while increasing average travel speed.

%Anjie Qiu 
%%security
\subsection{Security Management}

In the domain of network security, the integration of AI and ML into 6G-V2X systems is revolutionizing network security, communication reliability, autonomous driving safety, and data privacy. These technologies address the complex challenges posed by the dynamic and heterogeneous nature of 6G networks, ensuring robust security frameworks and reliable communication systems.  Asha et al. \cite{AJ:S_Asha_2024}, propose a FL-based network slicing system to dynamically allocate network resources while maintaining privacy, achieving a 94.2\% privacy protection score. Patel et al. \cite{AJ:patel2022}, introduce a Blockchain-based architecture and integrate FL to store and verify local model updates, ensuring the integrity and reliability of the learning process. Nair et al. \cite{AJ:Nair2022}, use AI-powered algorithms to classify normal and malicious traffic in V2X networks, achieving up to 97\% accuracy against network intrusions. It uses AI-empowered secure communication frameworks, combining blockchain and ML to encrypt and authenticate V2X messages in the 6G Network.
% \par
% In the domain of communication reliability in 6G-V2X, Patel et al. \cite{AJ:patel2022} leverages edge computing in 6G networks to reduce latency and improve communication efficiency by applying FL architecture among real-time V2X transmission. Asha et al. \cite{AJ:S_Asha_2024}, uses Quantum-enhanced security methods like Quantum Key Distribution to integrate AI-driven traffic control systems to improve the reliability of communication and safety of autonomous driving in 6G.
\par
In the domain of autonomous driving, security is the key concern in the 6G environment. Raja et al.\cite{AJ:raja2024} propose to apply AI and ML to enhance security in 6G-V2X by verifying vehicle identities and safeguarding against unauthorized access. Osorio et al. \cite{AJ:Diana_Pamela_Moya_Osorio_2022}, emphasized the need for attribute authentication in Internet of Vehicles(IoV), where OBUs, RSUs, vehicles, and nomadic devices should be able to provide proof of being authorized owners of legitimate identifications, ensuring secure group communications and legitimate membership verification within the IoV ecosystems.
\par
In the domain of data privacy in 6G-V2X, FL is the most favored ML architecture. Hangdong et al.\cite{AJ:Kuang_Hangdong_2023}, propose the FedVPS scheme to improve communication efficiency and model prediction accuracy compared to the traditional FedAvg method, making it more effective for distributed ML in IoV. Wang et al.\cite{AJ:Di_Wang_2024}, systematically reviews the current technical means and research progress addressing cybersecurity threats and privacy protection challenges faced by autonomous vehicles, highlighting key technical measures such as data encryption, access control and intrusion detection to enhance overall V2X system security.

\begin{table*}[ht]
\caption{AI Models Applied in 6G-V2X Applications}
\centering
\begin{tabular}{|>{\centering\arraybackslash}m{8cm}|l|l|}
\hline
\textbf{Representative Applications in 6G-V2X communication} & \textbf{AI Models} & \textbf{Ref.} \\
\hline

\multirow{3}{*}{} 
   Intelligent Resource Management & DRL & \cite{ji2024rlra,khan2024rlra} \\
    \cline{2-3}
    & LLM & \cite{liu2024llm} \\
    \cline{2-3}
    & FL  & \cite{bhardwaj2024flrl,prathiba2021FL} \\
\hline

\multirow{2}{*}{}
   AI-Powered Beamforming in Millimeter-Wave  (mmWave) \& Terahertz (THz) Bands & FL & \cite{bhardwaj2023flbf,bhardwaj2024flbf} \\
    \cline{2-3}
    & DL & \cite{moon2020dlbf} \\
\hline

\multirow{2}{*}{} 
   Intelligent Traffic Management & ML  & \cite{khalid2025modernizing,qiu2024advanced} \\
    \cline{2-3}
    & DRL & \cite{zhou2024intelligent} \\
\hline

\multirow{2}{*}{} 
    Security Management & FL & \cite{AJ:S_Asha_2024,AJ:Kuang_Hangdong_2023} \\
    \cline{2-3}
    & ML & \cite{AJ:Diana_Pamela_Moya_Osorio_2022} \\
\hline

\end{tabular}
\label{tab:6g-v2x-ai}
\end{table*}

\section{Challenges of AI for 6G-V2X}

Integrating AI into 6G-V2X systems introduces a range of complex challenges that must be addressed to ensure effective deployment and performance. 

One of the primary concerns is achieving real-time processing, as many 6G-V2X applications require ultra-low latency and high reliability. Traditional AI models may struggle to meet these strict timing requirements, particularly when operating on resource-constrained edge devices like RSUs.

Data privacy and security present another significant challenge \cite{osorio2022towards}. AI systems often require large volumes of data to train effectively, but sharing data between vehicles, infrastructure, and centralized servers raises concerns over user privacy and potential vulnerabilities to cyberattacks or adversarial manipulation. Moreover, the data collected in V2X environments is often limited, heterogeneous, and non-Independent and Identically Distributed (non-IID), which affects the ability of AI models to generalize across different driving scenarios and regions.

Resource limitations are also a critical issue. Vehicles and RSUs typically have constrained computational power and memory, making it difficult to deploy and update large AI models on the edge \cite{rajalakshmi2024towards}. In distributed learning approaches, such as FL, communication overhead can further hinder performance, especially in high-mobility environments where network conditions frequently change.

Scalability and robustness are essential for AI systems to adapt to dynamic traffic conditions, diverse user behaviors, and environmental changes. Ensuring that AI models remain reliable and effective in uncertain or unexpected situations, such as sensor failures, low visibility, or infrastructure disruption,s is a major technical hurdle. Finally, the absence of unified standards for AI integration in V2X networks impedes interoperability and consistent implementation across different manufacturers, regions, and regulatory frameworks.

Overall, addressing these challenges is crucial for realizing the full potential of AI in 6G-V2X, enabling safe, efficient, and intelligent transportation systems of the future.

\section{conclusion}

In this paper, we provide an overview of the latest advancements in AI techniques, including ML, DL, RL, LLMs, and FL, as they pertain to 6G-V2X communications. We also conduct a thorough review of recent research that leverages these advanced AI models to address key challenges and enable representative applications in the 6G-V2X domain, such as intelligent resource allocation, beamforming, intelligent traffic management, and Security. AI integrated for 6G-V2X communication is the future mobility revolution. Integrating AI with 6G-V2X will make transportation safer, smarter, and more efficient. AI will enable real-time decision-making for autonomous driving; Optimize traffic flow and reduce congestion; Enhance emergency response and accident prevention; Personalize in-vehicle experiences. This article aims to offer valuable insights into 6G-V2X communication, highlighting the integration of AI technologies to inspire further research and innovation in the practical design, testing, and deployment of these next-generation intelligent transportation systems.

\section{acknowledgement}
This work has been supported by the Federal Ministry of Education and Research of the Federal Republic of Germany (BMBF) as part of the Open6GHub project with funding number 16KISK004. The authors would like to express their appreciation for the contributions of all Open6GHub partners. The authors alone are responsible for the content of the paper, which does not necessarily represent the project.

\bibliographystyle{IEEEtran}
\bibliography{references}

\begin{thebibliography}{10}
\providecommand{\url}[1]{#1}
\csname url@samestyle\endcsname
\providecommand{\newblock}{\relax}
\providecommand{\bibinfo}[2]{#2}
\providecommand{\BIBentrySTDinterwordspacing}{\spaceskip=0pt\relax}
\providecommand{\BIBentryALTinterwordstretchfactor}{4}
\providecommand{\BIBentryALTinterwordspacing}{\spaceskip=\fontdimen2\font plus
\BIBentryALTinterwordstretchfactor\fontdimen3\font minus \fontdimen4\font\relax}
\providecommand{\BIBforeignlanguage}[2]{{%
\expandafter\ifx\csname l@#1\endcsname\relax
\typeout{** WARNING: IEEEtran.bst: No hyphenation pattern has been}%
\typeout{** loaded for the language `#1'. Using the pattern for}%
\typeout{** the default language instead.}%
\else
\language=\csname l@#1\endcsname
\fi
#2}}
\providecommand{\BIBdecl}{\relax}
\BIBdecl

\bibitem{noor20226g}
M.~Noor-A-Rahim, Z.~Liu, H.~Lee, M.~O. Khyam, J.~He, D.~Pesch, K.~Moessner, W.~Saad, and H.~V. Poor, ``6g for vehicle-to-everything (v2x) communications: Enabling technologies, challenges, and opportunities,'' \emph{Proceedings of the IEEE}, 2022.

\bibitem{3GP15}
``Study on new radio (nr) to support non-terrestrial networks,'' 3rd Generation Partnership Project (3GPP), Tech. Rep. 38.811, March 2022, version 15.4.0, Release 15.

\bibitem{IEEE07}
S.~Eichler, ``Performance evaluation of the ieee 802.11 p wave communication standard,'' in \emph{2007 IEEE 66th Vehicular Technology Conference}.\hskip 1em plus 0.5em minus 0.4em\relax IEEE, 2007, pp. 2199--2203.

\bibitem{3GP16}
\BIBentryALTinterwordspacing
``Study on {LTE}-based vehicle-to-everything ({V2X}) services,'' 3rd Generation Partnership Project (3GPP), Tech. Rep. 36.885, jun 2016, version 14.0.0, Release 14. [Online]. Available: \url{https://portal.3gpp.org/desktopmodules/Specifications /SpecificationDetails.aspx?specificationId=2934}
\BIBentrySTDinterwordspacing

\bibitem{3GP19}
\BIBentryALTinterwordspacing
``Study on {NR} vehicle-to-everything ({V2X}); ({R}elease 16),'' 3rd Generation Partnership Project (3GPP), Tech. Rep. 38.885, Jun 2019. [Online]. Available: \url{https://portal.3gpp.org/desktopmodules/Specifications/ SpecificationDetails.aspx?specificationId=3498}
\BIBentrySTDinterwordspacing

\bibitem{3GP21}
\BIBentryALTinterwordspacing
``Summary for {RAN} {Rel-18} package,'' 3rd Generation Partnership Project (3GPP), Tech. Rep. RP-213468, dec 2021, release 18. [Online]. Available: \url{\url{https://www.3gpp.org/specifications-technologies/releases/release-18}}
\BIBentrySTDinterwordspacing

\bibitem{nokia2023}
\BIBentryALTinterwordspacing
N.~W. Paper, ``Toward a 6g ai-native air interface,'' Online, 2023, accessed October 2023. [Online]. Available: \url{\url{https://onestore.nokia.com/asset/210299}}
\BIBentrySTDinterwordspacing

\bibitem{6gflagship}
\BIBentryALTinterwordspacing
G.~Flagship, ``6g research visions white paper series,'' Online, 2023, accessed October 2023. [Online]. Available: \url{\url{https://www.6gflagship.com/white-papers/}}
\BIBentrySTDinterwordspacing

\bibitem{sanghvi2021res6edge}
J.~Sanghvi, P.~Bhattacharya, S.~Tanwar, R.~Gupta, N.~Kumar, and M.~Guizani, ``Res6edge: An edge-ai enabled resource sharing scheme for c-v2x communications towards 6g,'' in \emph{2021 International Wireless Communications and Mobile Computing (IWCMC)}.\hskip 1em plus 0.5em minus 0.4em\relax IEEE, 2021, pp. 149--154.

\bibitem{tong2019ai}
W.~Tong, A.~Hussain, W.~X. Bo, and S.~Maharjan, ``Artificial intelligence for vehicle-to-everything: A survey,'' \emph{IEEE Access}, vol.~7, pp. 10\,823--10\,843, 2019.

\bibitem{christopoulou2023artificial}
M.~Christopoulou, S.~Barmpounakis, H.~Koumaras, and A.~Kaloxylos, ``Artificial intelligence and machine learning as key enablers for v2x communications: A comprehensive survey,'' \emph{Vehicular Communications}, vol.~39, p. 100569, 2023.

\bibitem{AJ:olayode2020}
\BIBentryALTinterwordspacing
O.~Olayode, L.~Tartibu, and M.~Okwu, ``Application of artificial intelligence in traffic control system of non-autonomous vehicles at signalized road intersection.'' \emph{Procedia CIRP}, vol.~91, pp. 194--200, 2020, enhancing design through the 4th Industrial Revolution Thinking. [Online]. Available: \url{https://www.sciencedirect.com/science/article/pii/S2212827120308076}
\BIBentrySTDinterwordspacing

\bibitem{AJ:omar2021}
A.~A. Omar, M.~M. Farag, and R.~A. Alhamad, ``Artifical intelligence: New paradigm in deep space exploration,'' in \emph{2021 14th International Conference on Developments in eSystems Engineering (DeSE)}, 2021, pp. 438--442.

\bibitem{ali2021machine}
E.~S. Ali, M.~K. Hasan, R.~Hassan, R.~A. Saeed, M.~B. Hassan, S.~Islam, N.~S. Nafi, and S.~Bevinakoppa, ``Machine learning technologies for secure vehicular communication in internet of vehicles: recent advances and applications,'' \emph{Security and Communication Networks}, vol. 2021, no.~1, p. 8868355, 2021.

\bibitem{gyawali2021deep}
S.~Gyawali, Y.~Qian, and R.~Q. Hu, ``Deep reinforcement learning based dynamic reputation policy in 5g based vehicular communication networks,'' \emph{IEEE Transactions on Vehicular Technology}, vol.~70, no.~6, pp. 6136--6146, 2021.

\bibitem{tan2022machine}
K.~Tan, D.~Bremner, J.~Le~Kernec, L.~Zhang, and M.~Imran, ``Machine learning in vehicular networking: An overview,'' \emph{Digital Communications and Networks}, vol.~8, no.~1, pp. 18--24, 2022.

\bibitem{aggarwal2018dl}
C.~C. Aggarwal \emph{et~al.}, \emph{Neural networks and deep learning}.\hskip 1em plus 0.5em minus 0.4em\relax Springer, 2018, vol.~10, no. 978.

\bibitem{yao1999ann}
X.~Yao, ``Evolving artificial neural networks,'' \emph{Proceedings of the IEEE}, vol.~87, no.~9, pp. 1423--1447, 1999.

\bibitem{marzuk2025deep}
F.~Marzuk, A.~Vejar, and P.~Cho{\l}da, ``Deep reinforcement learning for energy-efficient 6g v2x networks,'' \emph{Electronics}, vol.~14, no.~6, p. 1148, 2025.

\bibitem{aradi2020survey}
S.~Aradi, ``Survey of deep reinforcement learning for motion planning of autonomous vehicles,'' \emph{IEEE Transactions on Intelligent Transportation Systems}, vol.~23, no.~2, pp. 740--759, 2020.

\bibitem{gui2021gans}
J.~Gui, Z.~Sun, Y.~Wen, D.~Tao, and J.~Ye, ``A review on generative adversarial networks: Algorithms, theory, and applications,'' \emph{IEEE transactions on knowledge and data engineering}, vol.~35, no.~4, pp. 3313--3332, 2021.

\bibitem{doersch2016tutorial}
C.~Doersch, ``Tutorial on variational autoencoders,'' \emph{arXiv preprint arXiv:1606.05908}, 2016.

\bibitem{zhang2024generative}
R.~Zhang, K.~Xiong, H.~Du, D.~Niyato, J.~Kang, X.~Shen, and H.~V. Poor, ``Generative ai-enabled vehicular networks: Fundamentals, framework, and case study,'' \emph{IEEE Network}, 2024.

\bibitem{zhao2023llm}
W.~X. Zhao, K.~Zhou, J.~Li, T.~Tang, X.~Wang, Y.~Hou, Y.~Min, B.~Zhang, J.~Zhang, Z.~Dong \emph{et~al.}, ``A survey of large language models,'' \emph{arXiv preprint arXiv:2303.18223}, vol.~1, no.~2, 2023.

\bibitem{AJ:Zhiyuan_Zhou_2024}
\BIBentryALTinterwordspacing
``Safedrive: Knowledge- and data-driven risk-sensitive decision-making for autonomous vehicles with large language models,'' 2024. [Online]. Available: \url{http://arxiv.org/pdf/2412.13238}
\BIBentrySTDinterwordspacing

\bibitem{achiam2023gpt}
J.~Achiam, S.~Adler, S.~Agarwal, L.~Ahmad, I.~Akkaya, F.~L. Aleman, D.~Almeida, J.~Altenschmidt, S.~Altman, S.~Anadkat \emph{et~al.}, ``Gpt-4 technical report,'' \emph{arXiv preprint arXiv:2303.08774}, 2023.

\bibitem{tataria20216g}
H.~Tataria, M.~Shafi, A.~F. Molisch, M.~Dohler, H.~Sj{\"o}land, and F.~Tufvesson, ``6g wireless systems: Vision, requirements, challenges, insights, and opportunities,'' \emph{Proceedings of the IEEE}, vol. 109, no.~7, pp. 1166--1199, 2021.

\bibitem{tong2024connectgpt}
K.~Tong and S.~Solmaz, ``Connectgpt: Connect large language models with connected and automated vehicles,'' in \emph{2024 IEEE Intelligent Vehicles Symposium (IV)}.\hskip 1em plus 0.5em minus 0.4em\relax IEEE, 2024, pp. 581--588.

\bibitem{panyam2025survey}
S.~Panyam, A.~Donvir, G.~Paliwal, and P.~Gujar, ``Survey of llms and ai agents in v2x: Simulation, analysis \& architectures,'' in \emph{2025 Systems of Signals Generating and Processing in the Field of on Board Communications}.\hskip 1em plus 0.5em minus 0.4em\relax IEEE, 2025, pp. 1--11.

\bibitem{samarakoon2019fl}
S.~Samarakoon, M.~Bennis, W.~Saad, and M.~Debbah, ``Distributed federated learning for ultra-reliable low-latency vehicular communications,'' \emph{IEEE Transactions on Communications}, vol.~68, no.~2, pp. 1146--1159, 2019.

\bibitem{AJ:Biagio_Boi_2024}
``Decentralized identity management and privacy-enhanced federated learning for automotive systems: A novel framework,'' 2024.

\bibitem{zhang2024survey}
R.~Zhang, J.~Mao, H.~Wang, B.~Li, X.~Cheng, and L.~Yang, ``A survey on federated learning in intelligent transportation systems,'' \emph{IEEE Transactions on Intelligent Vehicles}, 2024.

\bibitem{he2017rl}
Y.~He, N.~Zhao, and H.~Yin, ``Integrated networking, caching, and computing for connected vehicles: A deep reinforcement learning approach,'' \emph{IEEE transactions on vehicular technology}, vol.~67, no.~1, pp. 44--55, 2017.

\bibitem{ji2024rlra}
M.~Ji, Q.~Wu, P.~Fan, N.~Cheng, W.~Chen, J.~Wang, and K.~B. Letaief, ``Graph neural networks and deep reinforcement learning based resource allocation for v2x communications,'' \emph{IEEE Internet of Things Journal}, 2024.

\bibitem{khan2024rlra}
N.~Khan and S.~Coleri, ``Event-triggered reinforcement learning based joint resource allocation for ultra-reliable low-latency v2x communications,'' \emph{IEEE Transactions on Vehicular Technology}, 2024.

\bibitem{liu2024llm}
C.~Liu and J.~Zhao, ``Resource allocation in large language model integrated 6g vehicular networks,'' in \emph{2024 IEEE 99th Vehicular Technology Conference (VTC2024-Spring)}.\hskip 1em plus 0.5em minus 0.4em\relax IEEE, 2024, pp. 1--6.

\bibitem{bhardwaj2024flrl}
S.~Bhardwaj, D.-H. Kim, and D.-S. Kim, ``Federated learning-based resource allocation for v2x communications,'' \emph{IEEE Transactions on Intelligent Transportation Systems}, 2024.

\bibitem{prathiba2021FL}
S.~B. Prathiba, G.~Raja, S.~Anbalagan, K.~Dev, S.~Gurumoorthy, and A.~P. Sankaran, ``Federated learning empowered computation offloading and resource management in 6g-v2x,'' \emph{IEEE Transactions on Network Science and Engineering}, vol.~9, no.~5, pp. 3234--3243, 2021.

\bibitem{mizmizi20216g}
M.~Mizmizi, M.~Brambilla, D.~Tagliaferri, C.~Mazzucco, M.~Debbah, T.~Mach, R.~Simeone, S.~Mandelli, V.~Frascolla, R.~Lombardi \emph{et~al.}, ``6g v2x technologies and orchestrated sensing for autonomous driving,'' \emph{arXiv preprint arXiv:2106.16146}, 2021.

\bibitem{forsythe2004mimo}
K.~Forsythe, D.~Bliss, and G.~Fawcett, ``Multiple-input multiple-output (mimo) radar: Performance issues,'' in \emph{Conference Record of the Thirty-Eighth Asilomar Conference on Signals, Systems and Computers, 2004.}, vol.~1.\hskip 1em plus 0.5em minus 0.4em\relax IEEE, 2004, pp. 310--315.

\bibitem{moon2020dlbf}
S.~Moon, H.~Kim, and I.~Hwang, ``Deep learning-based channel estimation and tracking for millimeter-wave vehicular communications,'' \emph{Journal of Communications and Networks}, vol.~22, no.~3, pp. 177--184, 2020.

\bibitem{bhardwaj2023flbf}
S.~Bhardwaj and D.-S. Kim, ``Federated learning-based joint radar-communication mmwave beamtracking with imperfect csi for v2x communications,'' in \emph{2023 Fourteenth International Conference on Ubiquitous and Future Networks (ICUFN)}.\hskip 1em plus 0.5em minus 0.4em\relax IEEE, 2023, pp. 201--206.

\bibitem{bhardwaj2024flbf}
S.~Bhardwaj and D.~S. Kim, ``Federated learning mm wave beamforming for v2x communications with imperfect csi and doppler shift,'' in \emph{2024 Fifteenth International Conference on Ubiquitous and Future Networks (ICUFN)}.\hskip 1em plus 0.5em minus 0.4em\relax IEEE, 2024, pp. 410--415.

\bibitem{khalid2025modernizing}
M.~Khalid, M.~Awais, C.~Jisi, M.~Ahmad, and B.-h. Roh, ``Modernizing transit: Intelligent traffic and transportation management with artificial intelligence in the era of 5g and 6g,'' \emph{The Intersection of 6G, AI/Machine Learning, and Embedded Systems: Pioneering Intelligent Wireless Technologies}, p. 208, 2025.

\bibitem{qiu2024advanced}
A.~Qiu, P.~A. Sathish, D.~Wang, and H.~D. Schotten, ``Advanced traffic demand generation in sumo: Ml-based prediction of flow rate based on real-world measured datasets,'' in \emph{2024 IEEE 99th Vehicular Technology Conference (VTC2024-Spring)}.\hskip 1em plus 0.5em minus 0.4em\relax IEEE, 2024, pp. 1--7.

\bibitem{zhou2024intelligent}
S.~Zhou, X.~Chen, C.~Li, W.~Chang, F.~Wei, and L.~Yang, ``Intelligent road network management supported by 6g and deep reinforcement learning,'' \emph{IEEE Transactions on Intelligent Transportation Systems}, 2024.

\bibitem{AJ:S_Asha_2024}
``Harnessing 6g technology for seamless autonomous vehicle communication and performance enhancement,'' pp. 99--104, 2024.

\bibitem{AJ:patel2022}
V.~A. Patel, P.~Bhattacharya, S.~Tanwar, N.~K. Jadav, and R.~Gupta, ``Bfledge: Blockchain based federated edge learning scheme in v2x underlying 6g communications,'' in \emph{2022 12th International Conference on Cloud Computing, Data Science \& Engineering (Confluence)}, 2022, pp. 146--152.

\bibitem{AJ:Nair2022}
A.~R. Nair, N.~K. Jadav, R.~Gupta, and S.~Tanwar, ``Ai-empowered secure data communication in v2x environment with 6g network,'' in \emph{IEEE INFOCOM 2022 - IEEE Conference on Computer Communications Workshops (INFOCOM WKSHPS)}, 2022, pp. 1--6.

\bibitem{AJ:raja2024}
D.~Raja, Z.~Abas, C.~Akula, Y.~Kumar, G.~Kumar, and V.~Eswari, ``Artificial intelligence powered internet of vehicles: securing connected vehicles in 6g,'' \emph{Indonesian Journal of Electrical Engineering and Computer Science}, vol.~35, p. 213, 07 2024.

\bibitem{AJ:Diana_Pamela_Moya_Osorio_2022}
\BIBentryALTinterwordspacing
``Towards 6g-enabled internet of vehicles: Security and privacy,'' \emph{IEEE open journal of the Communications Society}, vol.~3, pp. 82--105, 2022. [Online]. Available: \url{https://ieeexplore.ieee.org/ielx7/8782661/9702748/09681822.pdf}
\BIBentrySTDinterwordspacing

\bibitem{AJ:Kuang_Hangdong_2023}
``Fedvps: Federated learning for privacy and security of internet of vehicles on non-iid data,'' 2023, pp. 178--183.

\bibitem{AJ:Di_Wang_2024}
``Research on network security and privacy protection technology of autonomous vehicle,'' \emph{International Journal of Computer Science and Information Technology}, 2024.

\bibitem{osorio2022towards}
D.~P.~M. Osorio, I.~Ahmad, J.~D.~V. S{\'a}nchez, A.~Gurtov, J.~Scholliers, M.~Kutila, and P.~Porambage, ``Towards 6g-enabled internet of vehicles: Security and privacy,'' \emph{IEEE Open Journal of the Communications Society}, vol.~3, pp. 82--105, 2022.

\bibitem{rajalakshmi2024towards}
P.~Rajalakshmi \emph{et~al.}, ``Towards 6g v2x sidelink: Survey of resource allocation-mathematical formulations, challenges, and proposed solutions,'' \emph{IEEE Open Journal of Vehicular Technology}, 2024.

\end{thebibliography}

\end{document}